\documentstyle[prb,preprint,aps,epsf]{revtex}
\draft

\def\AmS{{\protect\the\textfont2
        A\kern-.1667em\lower.5ex\hbox{M}\kern-.125emS}}

\makeatletter
\def\thepage{1-\@arabic\c@page}
\def\@pnumwidth{2em}
\makeatother 
   \def\<{\langle}
   \def\>{\rangle}
\begin{document}

\title{Backscattering in carbon nanotubes : role of
quantum interference effects}
\author{Stephan Roche${\
}^{\ddagger}$, Fran\c cois
Triozon${\ }^{\dagger}$ and Angel Rubio${\ }^{*}$}
\address{ ${\ }^{\ddagger}$Commissariat \`a l'\'Energie
Atomique, DRFMC/SPSMS, Grenoble, France.\\
   ${\ }^{\dagger}$ LEPES-CNRS, avenue des Martyrs BP166,
38042 Grenoble, France.\\
${\ }^{*}$Departamento de F{\'\i}sica de Materiales, Facultad de
Ciencias Qu{\'\i}micas, Universidad del Pais Vasco/Euskal Herriko
Unibertsitatea, Apdo. $1072, 20018$ San Sebasti\'an/Donostia, Basque
Country, Spain.\\}
\maketitle

\begin{abstract}

  The backscattering contribution to the conductivity,
irrelevant for
metallic single-walled carbon nanotubes, is proved to
become much significant for
doped semiconducting systems, in agreement with
experiments. In the case of
multi-walled nanotubes, the intershell coupling is
further
shown to enhance the contribution of backscattering
for "metallic"
double-walled, whereas it remains insignificant for
"metallic/semiconducting"
double-walled systems. 
\end{abstract}

\vspace{0.2in}

\vfill\eject
Single-walled carbon nanotubes (SWNTs) can be either
metallic of semiconducting depending on their
helicities, i.e. how the graphene
sheet is rolled up\cite{Saito}. 
For weak uniform disorder, the application of the
Fermi golden rule for {\it metallic SWNTs} have
demonstrated $\mu m$ long mean free paths
\cite{Todorov} in agreement with experiments,
clearly pointing towards ballistic
transport\cite{CN-FET,Bachtold2,CN-FET2}.

Notwithstanding, quantum transport in SWNTs is richer
than in
the one-dimensional chain, given the implication of
 additional symmetries of electronic 
 eigenstates associated to the circumferential
helicity. 
 This has been widely illustrated through the
theoretical study of 
 conduction upon introduction of single defects such
as vacancies, 
 impurities or topological defects\cite{Disorder}. In particular, 
 the absence of backscattering was demonstrated for single
impurity with long range 
 potential in metallic tubes\cite{Saitobc} 
 and stepwise reduction of conductance was inferred from
multiple
scattering on a few lattice
impurities\cite{Saitobc,Choi}. Resonant electronic scattering by
defects was recently confirmed experimentally\cite{Lieber}.

On the other hand, electrostatically or chemically 
{\it doped semiconducting SWNTs} have 
been reported to behave as diffusive conductors with
short
mean free paths, several orders of magnitude
lower than the ones of reported structurally equivalent metallic
SWNTs\cite{CN-FET2,Louie,Kruger}. In all these experiments, the
mean free path ($l_{e}$) is deduced from the measured
conductance using $G\sim (e^{2}/h)\ell_{e}/L_{tube}$ 
($L_{tube}$ the length of the SWNT) and values range
from $2nm$\cite{Louie} to 
about $\sim 30$nm\cite{CN-FET2}. Upon
doping, 
the position of the chemical potential (Fermi 
energy) with respect to the charge neutrality point is
shifted 
downward (hole-doping) or upward (electron-doping) and hence may come closer
to a Van-Hove singularity. This may result in a factor
 $\sqrt{\<v^{2}\>}$ much smaller than the typical
$v_{F}$ deduced from the metallic SWNTs. Moreover, quantum interference
effects (QIE) responsable for localization in 1D-systems need to be clarified
in the context of carbon nanotubes. Indeed, on
multi-walled carbon nanotubes, Bachtold et al.\cite{AB-NT2} have reported
negative magnetoresistance and Aharonov-Bohm oscillations, consistent with the
manifestation of quantum interferences in the weak
localization regime. This experiment was interpreted by assuming a
current predominantly carried in the outermost shell (taken metallic), and 
mean free path and coherence lengths were deduced from conventional theory. It
is surprising that quantum interferences that have been described for the
two-dimensional propagation (with {\it many more conducting channels}), still
account properly for the behavior of a single metallic nanotube shell,
which only presents {\it two conducting channels} at the charge neutrality
point. In that perspective, the debate of ballistic \cite{MWNT-QQ} against
diffusive\cite{AB-NT2,MWNT-diff} conduction in MWNTs is a great issue of
concern.

All these considerations can be addressed by rewriting
 the conductance as
$(e^{2}\sqrt{\<v^2\>}\tau_{e}- \mid\delta\sigma\mid)/L_{tube}$
($\tau_{e}$ the mean free time
of eigenstates given by the Fermi
golden rule), and by evaluating properly the contribution of
backscattering. This quantum correction ($\mid \delta\sigma\mid$)
to the Bloch-Boltzmann conductivity ($\sigma_{BB}$), is related to
 the probability of return to the origin of electronic wavepackets, that is
connected to the Participation Ratio (PR), an
energy dependent quantity which measures the
"spreading" of the electron eigenstates commonly used to address
QIE in weak or strong localization
regimes\cite{Wegner,Carini}. For an eigenstate $\psi(E)$
described by its $N$ coefficients $\psi_i(E)$ in a tight-binding basis set,
the PRs read 

$$PR(E)={\displaystyle
({\displaystyle \sum_{i}^{N}}
\mid\psi_{i}(E)\mid^{2})^{2} / 
{\displaystyle
\sum_{i}^{N}}\mid\psi_{i}(E)\mid^{4}}$$

\noindent
and it can be shown that an average of the probability of return to the origin
in real space is equivalent to an average of the inverse PR on the
spectral bandwidth. Accordingly, the amplitude of quantum correction to
the electronic conductivity can be estimated as \cite{QIE}

$$\delta\sigma/\sigma_{BB}\sim
PR^{-1} (\hbox{for}\ N\to\infty)$$

\noindent
Eigenstates characterized by a linear
scaling in $N$ are uniformly extended and associated
with a vanishing
contribution of QIE, i.e. $\delta\sigma/\sigma_{BB}\to
0$. Instead, localized states are related to strong contributions of QIE, i.e.
$\delta\sigma/\sigma_{BB}\simeq 1$, whereas scaling laws $PR(N)=
N^{\alpha}$, with $0<\alpha<1$, indicate the relative strength of QIE.

In this letter we present detailed calculations of the PRs for different
tubes, addressing the role of disorder and intertube coupling in the
transport properties of nanotubes. We use the standard and reliable
one-electron tight binding model including intertube interactions (for MWNTs)
fitted to ab-initio calculations\cite{Saito2}. Disorder is included by a
random modulation of onsite energies within the range $[-W/2,W/2]$
($W=0.054,0.135,0.98eV$) that simulate chemical substitutions. The mean free
path associated for a given disorder is deduced from
$\ell_{e}\sim (\gamma_{0}/W)^{2}d_{nt}$, ($d_{nt}$ is the nanotube
diameter, $\gamma_{0}=2.67eV$ the hopping between carbon
sites)\cite{Todorov,Mahan}.

{\it Effect of disorder in SWNTs}.-The density of
states (DoS) together with
the PRs of several metallic chiral and achiral SWNTs
are reported on Fig.1 (results are nearly identical for $W=0.054eV$
and $0.135eV$). For Fermi energies at the charge neutrality point, $PR=N$
for armchair and achiral tubes, whereas $PR\sim 2N/3$ for zig-zag
SWNTs. An
energy dependence relation between the position of
Van-Hove singularities
and the amplitude of PRs is also found. For the Bloch
states of zigzag and
armchair tubes (the central sub-band of armchairs
excepted), there is a
degeneracy due to mirror-inversion symmetry. Each
value of the wavevector is
associated to two Bloch states with the same energy,
but with an opposite
phase variation along the circumference~: 
$\psi_{n}^{+} =
\exp(ik\theta_{n}/2\pi )$ and $\psi_{n}^{-} = 
\exp(-ik\theta_{n}/2\pi )$,
where $k > 0$ is a positive integer and  $\theta_{n}$
is the polar angle of
the site $n$ located on a given  ring. By a linear
combination of $\psi^{+}$ and
$\psi^{-}$, a Bloch state with any  PR value between
$2N/3$ and $N$ can be
constructed. The lowest value is given by the
combination
$(\psi^{+}+\psi^{-})/\sqrt{2}$, which is a ``standing
wave'' along the
circumference~:  $\psi_{n}^{+}+\psi_{n}^{-} =
2\cos(k\theta_{n}/2\pi )$,
leading to  $PR = N(\< \cos^{2}(k\theta_{n}/2\pi) \>
)^{2}/ 
\< \cos^{4}(k\theta_{n}/2\pi) \> = 2N/3$. Due to this
uncertainty, 
the PR is thus not a well-defined quantity for
degenerate eigenstates. But a
small amount of disorder is enough to split degeneracy
and the PRs shown in
Fig.1  become meaningful. For nondegenerate
eigenstates (as found
in chiral metallic SWNTs and armchair SWNTs close to
Fermi energy),
perturbation theory applies, so that the PRs are not
much reduced by  disorder
and remain close to $N$.  Instead, for degenerate
states (zig-zag metallic
SWNTs), disorder favors standing waves, i.e. states
with a real wavefunction,
because the hamiltonian is real and symmetric. 
Thereby the PRs are close to
$2N/3$, which explains the general behavior shown in
Fig.1. Values
much smaller than $2N/3$  are attributed to standing
waves along the tube axis
$z$, obtained by  mixing $k_{z}$ and $-k_{z}$ Bloch
states close to the Van-Hove singularities. At the
charge neutrality point of the
metallic SWNT $(6,6)$, the
eigenstates are basically
insensitive to disorder and follow the linear scaling
expected for fully
extended states (inset Fig.2). This confirms that the
effect of small disorder is
purely marginal in metallic SWNTs close to Fermi
energy\cite{Todorov,Saitobc}.

On the contrary, a stronger contribution of QIE for doped
semiconducting SWNTs is demonstrated by the scaling behavior of PRs as
reported on Fig.2 (main frame). The chemical potential of the
(7,5) semiconducting tube has been upshifted by assuming a typical
dopant concentration of a few percent of carbon sites (within the rigid band
approximation). Departure from linear scaling is obtained ($PR=N^{\alpha}$)
with an increasing contribution of QIE with disorder strength ($\alpha\sim
0.98, 0.95$ for $W=0.054eV$ and $0.136eV$ respectively). For a larger disorder
strength ($W=0.98eV$), that corresponds to a mean free path $\ell_{e}\simeq
20nm$, the saturation of the PR provides an approximate localization length of
$\xi\simeq 40nm$. From the Thouless argument, it is believed
that $\xi\sim 4\times\ell_{e}$ for a metallic shell that provides two
conducting channels at the charge neutrality point\cite{Todorov,Mahan}. Hence,
for doped semiconducting tubes, the mixing between quantum channels induced by
substitutional disorder, results in an enhanced contribution of QIE, and
lower localization lengths. This demonstrates that the conduction mechanism in
the outermost shell of a MWNT depends on the position of the chemical
potential\cite{AB-NT2,MWNT-QQ}. To complete this argument, we add the effect 
of intertube coupling as it promotes charge transfer between shells in MWNTs.


{\it Effect of disorder and intershell coupling in
commensurate MWNTs}.-The characteristic
sensitivity to disorder of QIE in MWNTs is illustrated on
Fig.3
for small disorder ($W=0.054eV$). For the
double-walled
"metallic" armchair tube $(6,6)@(11,11)$, the average
PR close to the charge neutrality point is
roughly half the value for the isolated armchair tube,
for same disorder
parameter. According to the previous discussion, the
$\mid\delta\sigma\mid$
contribution is now not negligible
and contribute to a reduction of the conductivity (confirmed by the
scaling analysis of PR-not shown here). Such effect
is similar in double-walled "metallic" zig-zag tube, but it is reduced
when the outer shell is semiconducting. It can be understood from the fact
that, at the charge neutrality point, the states of such MWNTs are mainly
weighted in the inner metallic shell, so less sensitive to the delocalization
induced by intershell coupling, and less affected by disorder as in
metallic SWNTs. Hence, MWNTs consisting of $\sim 2/3$
of semiconducting shells remain {\it long ballistic
conductors at charge neutrality point}\cite{MWNT-QQ}.

   \noindent Acknowledgments:
    Financial support from {\small NAMITECH [ERBFMRX-CT96-0067(DG12-MITH)],
DGES (PB98-0345), COMELCAN( HPRN-CT-2000-00128), JCyL (VA28/99)} and {\small
C$^4$} are acknowledged.

\vfill\eject

\vfill\eject
\noindent 
{\bf Figure captions}:

\vspace{20pt}
\noindent
{\bf Figure 1}: PRs (solid lines) and TDoS (dashed lines)
for metallic armchair $(6,6)$, zigzag 
   $(18,0)$ and chiral $(9,6)$ tubes. PRs are
normalized to the number $N$ of 
   atoms, and have been averaged over a few disorder
configurations ($W=0.054eV$). TDoS are given in arbitrary units.

\vspace{20pt}
\noindent
{\bf Figure 2}: Comparison of the scaling of the PRs for 
metallic (6,6) (inset) and doped semiconducting (7,5) single-walled nanotube
(for which the chemical potential lies nearby a Van-Hove singularity,
$E_{F}/\gamma_{0}\sim 0.333$ upshifted with respect to the charge neutrality
point). The dashed lines indicate the linear scaling ($PR=N$). Comparison is
made for two values of disorder strength $W=0.054eV$(open circles) and
$0.136eV$(filled circles). The third disorder strength $W=0.98eV$(filled
diamonds) taken for (7,5) at the same Fermi energy leads to $\ell_{e}\sim
20nm$.

\vspace{20pt}
\noindent
{\bf Figure 3}: PRs for different double-walled tubes : "metallic" armchair
$(6,6)@(11,11)$, zigzag $(9,0)@(18,0)$, and "metallic/semiconducting" 
chiral $(9,6)@(15,10)$ nanotubes.

\vfill\eject

 \begin{figure}[htbp]
   \epsfxsize=15cm
   \centerline{\epsffile{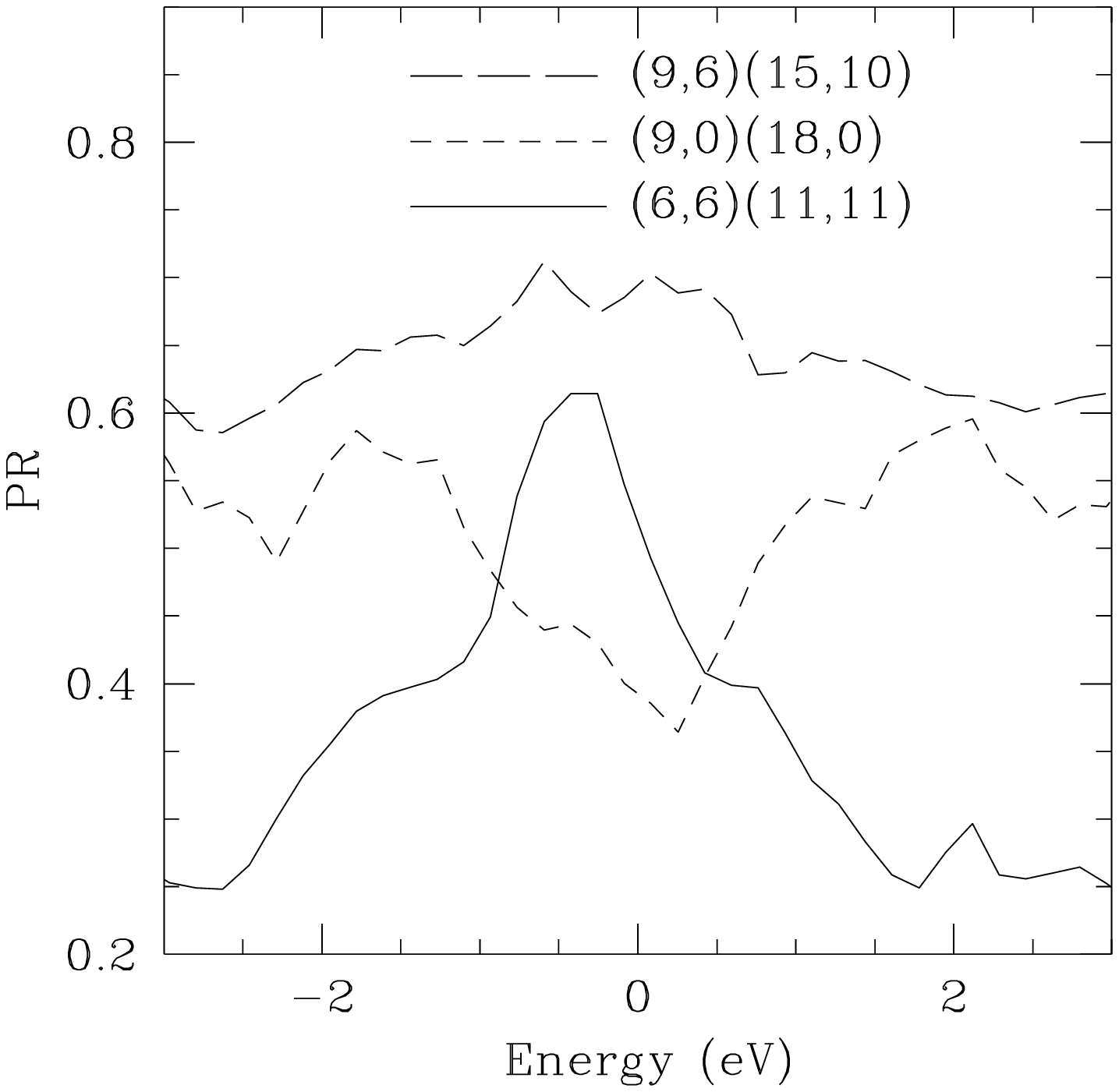}}

   \end{figure}

\vfill\eject

\begin{figure}[htbp]
   \epsfxsize=15cm
   \centerline{\epsffile{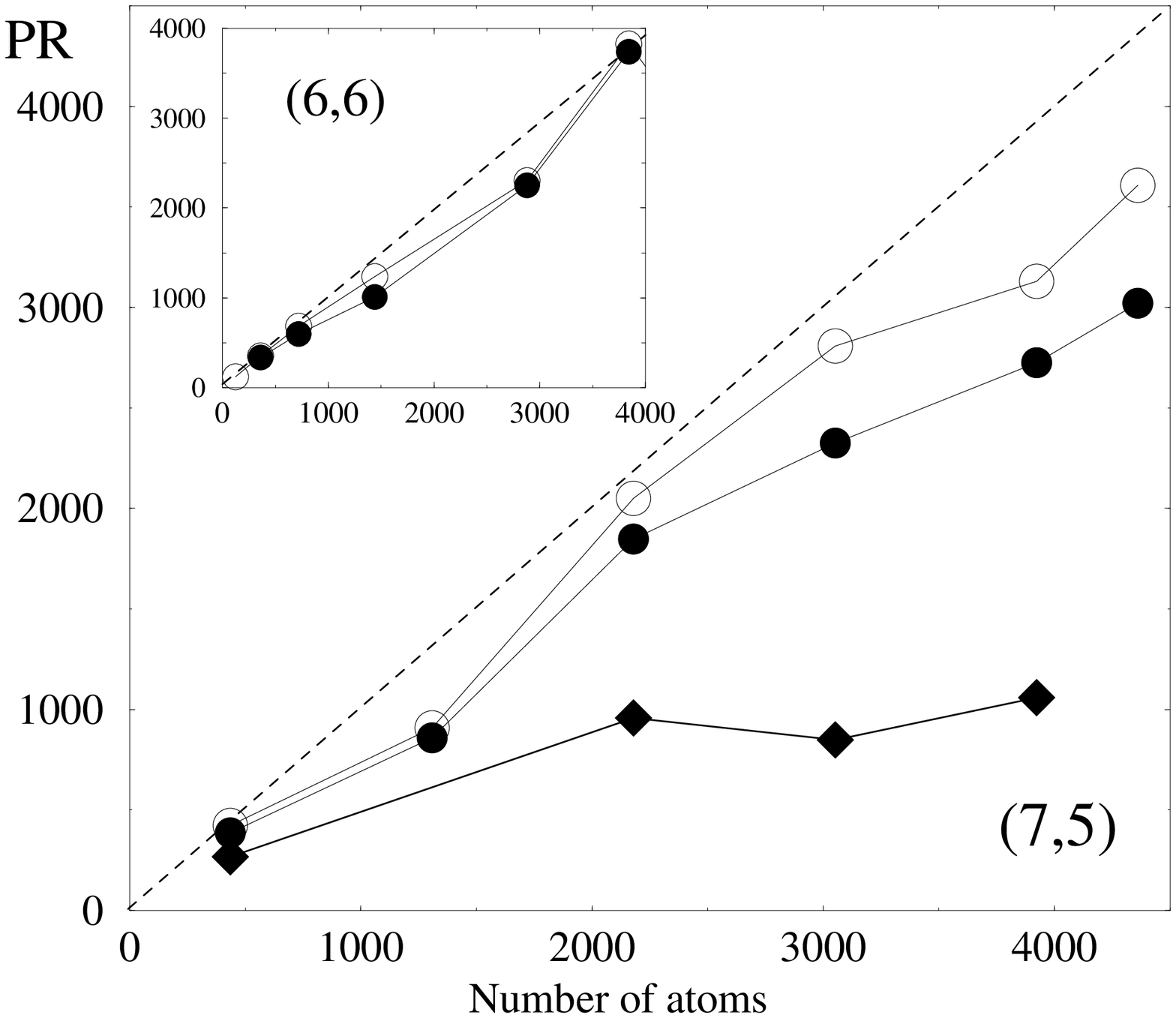}}         
      
\end{figure}

\vfill\eject

\begin{figure}[htbp]
   \epsfxsize=15cm
   \centerline{\epsffile{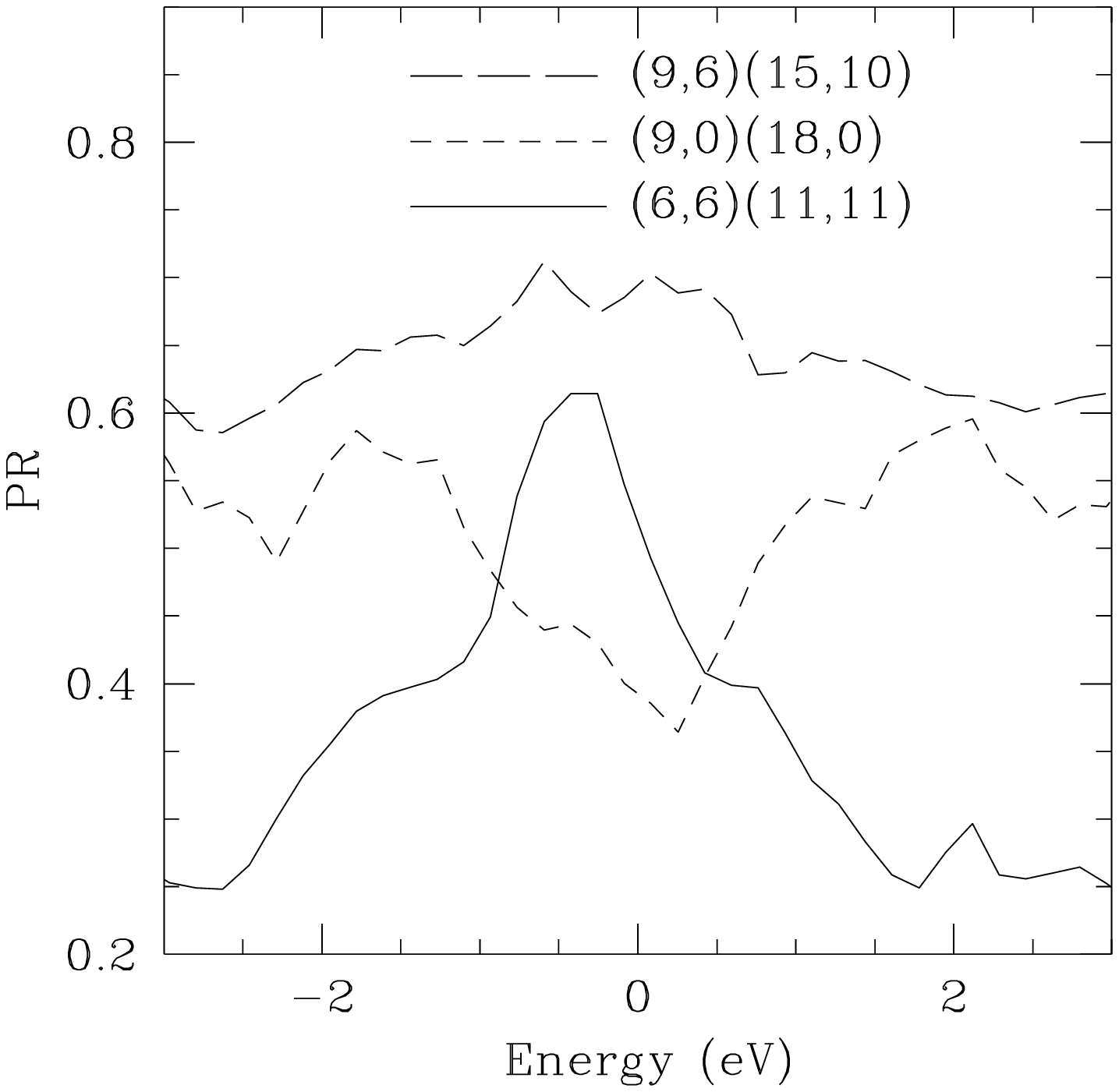}}
       
\end{figure}

\vfill\eject


\begin{thebibliography}{100}
\bibitem{QIE}{\small
$\mid \delta\sigma\mid$ is related to $\frac{2e^{2}}{h}2D\int_{0}^{\infty}dt
{\cal P}(r=0,t) e^{-t/\tau_{\varphi}}(1-e^{-t/\tau_{e}})$ \cite{Mesotr}, with
$D=v_{F}^{2}\tau_{e}$, $\tau_{e}=\ell_{e}/v_{F}$ the mean
free time, and $\tau_{\varphi}$ the inelastic time which limitates the
contribution of the probability of return to the origin ${\cal P}(r=0,t)$. Such
quantity is given by {\small $\mid\langle r\mid exp(-iHt/\hbar)\mid
r\rangle\mid^{2}$}.}
\end{thebibliography}
   \end{document}